\documentclass{article}

\usepackage{arxiv}

\usepackage[utf8]{inputenc} 
\usepackage[T1]{fontenc}    
\usepackage{hyperref}       
\usepackage{url}            
\usepackage{booktabs}       
\usepackage{amsfonts}       
\usepackage{nicefrac}       
\usepackage{microtype}      
\usepackage{lipsum}		
\usepackage{graphicx}
\usepackage{natbib}
\usepackage{doi}

\usepackage{multirow}
\usepackage{amsmath}
\usepackage{nomencl}

\title{Parametric Dynamic Mode Decomposition with multi-linear interpolation for prediction of thermal fields of Al$_2$O$_3$-water nanofluid flows at unseen parameters}

\author{ \href{https://orcid.org/0000-0003-1735-1918}{\includegraphics[scale=0.06]{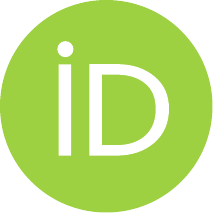}\hspace{1mm}Abhijith M S}\\
	Amrita School of Artificial Intelligence\\
        Amrita Vishwa Vidyapeetham\\
        Coimbatore\\
        India\\
	\texttt{ms\_abhijith@cb.amrita.edu} \\
	\And
	Sandra S\\
	Amrita School of Artificial Intelligence\\
        Amrita Vishwa Vidyapeetham\\
        Coimbatore\\
        India\\
	\texttt{s\_sandra1@cb.students.amrita.edu} \\
}

\renewcommand{\shorttitle}{Parametric DMD with MLI for prediction of thermal fields at unseen parameters}

\hypersetup{
pdftitle={manuscript_msa},
pdfsubject={Machine learning, CFD},
pdfauthor={Abhijith M S, Sandra S},
pdfkeywords={nanofluid, machine learning, More},
}

\begin{document}
\maketitle

\begin{abstract}
The study proposes a data-driven model which combines the Dynamic Mode Decomposition with multi-linear interpolation to predict the thermal fields of nanofluid flows at unseen Reynolds numbers (Re) and particle volume concentrations ($\epsilon$). The flow, considered for the study, is laminar and incompressible. The study employs an in-house Fortran-based solver to predict the thermal fields of Al$_2$O$_3$-water nanofluid flow through a two-dimensional rectangular channel, with the bottom wall subjected to a uniform heat flux. The performance of two models operating in one- and two-dimensional parametric spaces are investigated. Initially, a DMD with linear interpolation (DMD-LI) based solver is used for prediction of temperature of the nanofluid at any Re $>$ 100.
The DMD-LI based model, predicts temperature fields with a maximum percentage difference of just 0.0273\%, in comparison with the CFD-based solver at Re =960, and $\epsilon$ = 1.0\%. The corresponding difference in the average Nusselt numbers is only 0.39\%. Following that a DMD with bi-linear interpolation (DMD-BLI) based solver is used for prediction of temperature of the nanofluid at any Re $>$ 100 and $\epsilon$ $>$ 0.5\%. The performance of two different ways of stacking the data are also examined. When compared to the CFD-based model, the DMD-BLI-based model predicts the temperature fields with a maximum percentage difference of 0.21 \%, at Re = 800 and $\epsilon$ = 1.35\%. And the corresponding percentage difference in the average Nusselt number prediction is only 6.08\%. All the results are reported in detail.  Along side the important conclusions, the future scope of the study is also listed. 
\end{abstract}

\keywords{nanofluid \and machine learning \and DMD \and linear interpolation \and parametric dynamical mode decomposition \and partial differential equations}

\section{Introduction}

Humanity's enduring curiosity about comprehending and exploring dynamical systems, stems from the desire to uncover the fundamental laws that govern the evolution of the universe. The research in the field of computational fluid dynamics often not only requires predicting the systems behaviour for future time instants, but also understanding their parametric dependencies to forecast behaviour for unseen parameters. The introduction of parameters expands the dimension of the problem space, necessitating high computational efficiency, to solve these equations across varying parametric values \cite{cao2024solving}.

We can mitigate this curse of dimensionality by using the technique of reduced order modelling (ROM) \cite{huhn2023parametric}, which employs mathematical approaches to derive smaller models without compromising solution accuracy. Several techniques fall under ROM, including reduced basis methods (RBM) \cite{quarteroni2015reduced, chakir2019non}, proper orthogonal decomposition (POD) \cite{zokagoa2012pod, xiao2012enhanced, xiao2015non, berkooz1993proper}, and proper generalised decomposition \cite{chinesta2011short}. 

Dynamic mode decomposition (DMD) \cite{schmid2010dynamic}, on the other hand, is a non intrusive method \cite{huhn2023parametric} that focuses on system dynamics. It processes high dimensional sequential data, captures and simplifies complex behaviours \cite{schmid2022dynamic},  including those in non-linear systems 
\cite{rozza2018advances},and identifies and represents the system as a linear combination of dominant structures within the data. Originally developed in the fluid dynamics community \cite{schmid2010dynamic}, DMD has expanded to offer a wide range of applicability across multiple domains, including robotics \cite{berger2015estimation}, financial modelling \cite{mann2016dynamic, unnithan2019}, blood flow physics \cite{habibi2020data}, image segmentation \cite{sikha2018}, image classification \cite{rahul2019dynamic},  DNA sequence classification \cite{aadhithya2023dmd} and more.


As already mentioned, here the intention is to predict the flow/ thermal behaviour at unseen parameters using DMD. There have been several attempts to mitigate the challenges associated with parameterized non linear dynamical systems, particularly through non-intrusive analysis \cite{yu2019non} that offer potential solutions. The literature on non-intrusive analysis of parameterized non-linear dynamical systems within fluid dynamics, focuses on three core aspects: surrogate modelling \cite{hoang2022projection, li2023deep, nikolopoulos2021non}, sampling strategy and reduced order modelling (ROM)\cite{hess2023data, duan2024non, sun2023parametric}.

An approach involving DMD and regression for predicting the snapshots of data at future time and unseen parameters is reported in \cite{andreuzzi2023}. They considered the data to involve high fidelity snapshots taken at uniform time intervals and across various parametric configurations. These snapshots are then undergone dimensionality reduction using POD and subsequently reorganized to serve as input for the DMD algorithm. DMD is applied to predict the reduced snapshots for future time steps. Finally, the parametric manifold is reconstructed using regression technique that approximates the relationship between the parameters and the reduced snapshots. They reported that for problems involving flow prediction employing Navier-Stokes equation, the dimensionality reduction leads to inadequate representation of the true state.

This study introduces a novel approach that combines Dynamic Mode Decomposition (DMD) with linear interpolation to predict the thermal fields of nanofluid flows. The flow under consideration is laminar and incompressible. To simplify the complexity of temporal data, only quasi-steady-state data for different parameters is used. Inorder, to avoid the limitations of reduced-order representations, such as those encountered in Proper Orthogonal Decomposition (POD) \cite{andreuzzi2023}, this study does not consider any dimensionality reduction. The study employs an in-house Fortran-based solver to predict the thermal fields of Al$_2$O$_3$-water nanofluid flow through a two-dimensional rectangular channel (referred to as the 'CFD-based' model), with the bottom wall subjected to a uniform heat flux. The simulation data is then utilized to develop a DMD with linear interpolation model capable of predicting thermal fields for nanofluids at previously unseen parameters. The intention here is to propose a DMD with multi-linear interpolation based models capable of predicting the thermal fields (and also the heat transfer performance) in a two dimensional channel with Al$_2$O$_3$-water nanofluid flow at different values of reynolds numbers and nanoparticle volume concentration.

Initially the governing equations used in the in-house solver are discussed in Section - \ref{ge}. Following that the core idea behind the DMD - multi-linear interpolation based model is discussed in Section - \ref{dmdli}. Prior to the results and discussions in the Section - \ref{result}, the grid independence and validation of the in-house solver are also presented (see sections \ref{gi} and \ref{valid} ).

\section{Governing Equations}\label{ge}

The in-house Fortran based solver considers a laminar, incompressible, unsteady, two-dimensional homogeneous model for the nanofluid flows. The governing equations for mass, momentum and energy can be given as shown below: 

\subsection{Continuity Equation}

\begin{equation}
    \nabla.{\bf v} = 0
\end{equation}

\subsection{Momentum Equation}

\begin{equation}
  \rho_{nf} \left( \frac{\partial {\bf v}}{ \partial t } + \left( {\bf v}.\nabla \right)  {\bf v} \right)  = - \nabla p +  \mu_{nf} \nabla.\left[ \left( \nabla {\bf v} + \nabla {\bf v}^T  \right) \right]  
\end{equation}

\subsection{Energy Equation}

\begin{equation}
\rho_{nf} c_{p, nf} \left( \frac{\partial \theta}{ \partial t } + \left( {\bf v}.\nabla \right)  {\theta} \right)  =  k_{nf} \nabla.\left( \nabla \theta \right)
\end{equation}

where, {\bf v} and $\theta$ represent the velocity (in m/s) vector  and temperature (in K) respectively. The thermo-physical properties of the nanofluid used in the above governing equations; such as, the density ($\rho_{nf}$, in kg/$m^3$), the specific heat capacity at constant pressure ($c_{p, nf}$ in J/kgK), and the thermal conductivity ($k_{nf}$ in W/mK) are expressed in terms of the nano particle concentration ($\epsilon$) and the individual properties of nano particles (denoted as `s') and the base fluid (denoted as $\ell$), as follows: 

\begin{eqnarray}
\rho_{nf} = (1-\epsilon)\rho_{\ell} + \epsilon \rho_s\\ 
\mu_{nf} = \frac{\mu_{\ell}}{\left( 1 - \epsilon \right)^{2.5}}\\
c_{p, nf} = \frac{1}{\rho_{nf}} \left((1-\epsilon)\rho_{\ell}c_{p,\ell} + \epsilon \rho_s c_{p,s} \right)
\end{eqnarray} 

\section{Dynamic Mode Decomposition with Linear Interpolation}\label{dmdli}

An issue with the conventional data-driven models is that the model performs satisfactorily only on those flow and thermal conditions considered in the training data. The flow and thermal conditions includes the boundary conditions, initial conditions, and parameters such as Reynolds number, nanoparticle concentration, heat flux, etc. Hence the prediction using the model on unseen parameters becomes a difficulty. An attempt is made in this study to tackle this limitation. The method of linear interpolation is used along with the dynamic decomposition in parametric space for the prediction on unseen parameters.

Using DMD, we seek a dynamic operator A, which satisfies;

\begin{equation}
    \text{\bf x}_{n+1} = \text{A} \text{\bf x}_n
    \label{xn+1=Axn}
\end{equation}

Here, {\bf x}$_{n+1}$ is the states of the system at instance 'n+1' in the parametric space. Also, we assume the instance `n+1' in parametric space (a K-dimensional space) has coordinates $\mathbb{\zeta}$ = ($\zeta_1$, $\zeta_2$, $\hdots$, $\zeta_K$), where $\zeta_k \in \mathbb{R}^{\text{M}_\text{k}}$ represents a distinct parameter, such as Reynolds number, particle concentration, or heat flux (also let, 1 $\leq$ k $\leq$ K). Similarly {\bf x}$_n$ represents the states at an instant `n'. The data in the form of `N' different snapshots {\bf x}, collected at equal intervals within the parametric space are stacked into two matrices, {\bf X} and {\bf X}$^\prime$ such that; 

\begin{eqnarray}
    \text{\bf X} = \left[ \text{ \bf x}_1, \text{ \bf x}_2, \hdots, \text{ \bf x}_{N-2},\text{ \bf x}_{N-1} \right]\\
    \text{\bf X}^ \prime = \left[ \text{ \bf x}_2, \text{ \bf x}_3, \hdots, \text{ \bf x}_{N-1},\text{ \bf x}_{N} \right]
\end{eqnarray}

where the total number of data instances (N) depends on the dimension of $\zeta_k$ (denoted as M$_k$), where 1 $\leq$ k $\leq$ K (note: `K' denotes the total number of parameters considered). Hence,

\begin{equation}
    N = M_1 \times M_2 \times M_3 \times \hdots \times M_K 
\end{equation}

Then, we seek a best fit operator A, using the DMD algorithm \cite{tu2013}, that fits the Eqn.\ref{Xprime=AX};
\begin{equation}
    \text{\bf X}^\prime \approx \text{A} \text{\bf X}
    \label{Xprime=AX}
\end{equation}

Now, for a future state prediction, say at instance `n+1' in the parametric space, we can write;

\begin{equation}
 \text{ \bf x}_{n+1} = \text{A} \text{ \bf x}_n = \text{A}\text{A} \text{ \bf x}_{n-1} = \text{A}^{n+1} \text{ \bf x}_0  
\end{equation}

For predicting a future state `$\hat{\text{\bf x}}$' at any set of K different parameters, $\hat{\mathbf{\zeta}}$ = ($\hat{\zeta_1}$, $\hat{\zeta_2}$, $\hdots$, $\hat{\zeta_K}$), we make use of multi linear interpolation in K-dimensional parametric space. Hence,

\begin{eqnarray}
\hat{\text{\bf x}} =  \sum_{i=1}^{2^K} {\text{\bf x}_i} w_i(\hat{\zeta}) 
\label{xhat_pred}
\end{eqnarray}

For the `i$^{th}$' vertex in the D-dimensional
unit hypercube, and w$_i$ ($\hat{\zeta}$) is the multilinear interpolation weight on the `i$^{th}$' unit hypercube vertex
$\xi_i$ $\in$ [0, 1]$^K$ taken in lexicographical order, computed from $\hat{\zeta}$ $\in$ [0, 1]$^K$ as; \cite{sharma2017, Zhang2021}

\begin{eqnarray}
w_i(\hat{\zeta}) = \prod_{k=1}^{K} \hat{\zeta}(k) ^{\xi_i(k)} \left( 1- \hat{\zeta}(k)^{1-\xi_i(k)} \right)
\end{eqnarray}

for all i = 1, 2, $\hdots$, 2$^K$.

A schematic representation of the linear interpolation in one and two dimensional parametric spaces are given in figures \ref{1D_b} and \ref{2D_b}. Note, here the states {\bf x}$_i$ (represented using black circles in figures \ref{1D_b} and \ref{2D_b}), mentioned in equation \ref{xhat_pred} are predicted using the DMD model. Then, the linear interpolation is employed to predict the required state $\hat{\text{\bf x}}$ using Eqn. \ref{xhat_pred}.

\section{Grid Independence Test} \label{gi}

\begin{center}
\begin{figure}
    \includegraphics[scale=0.89]{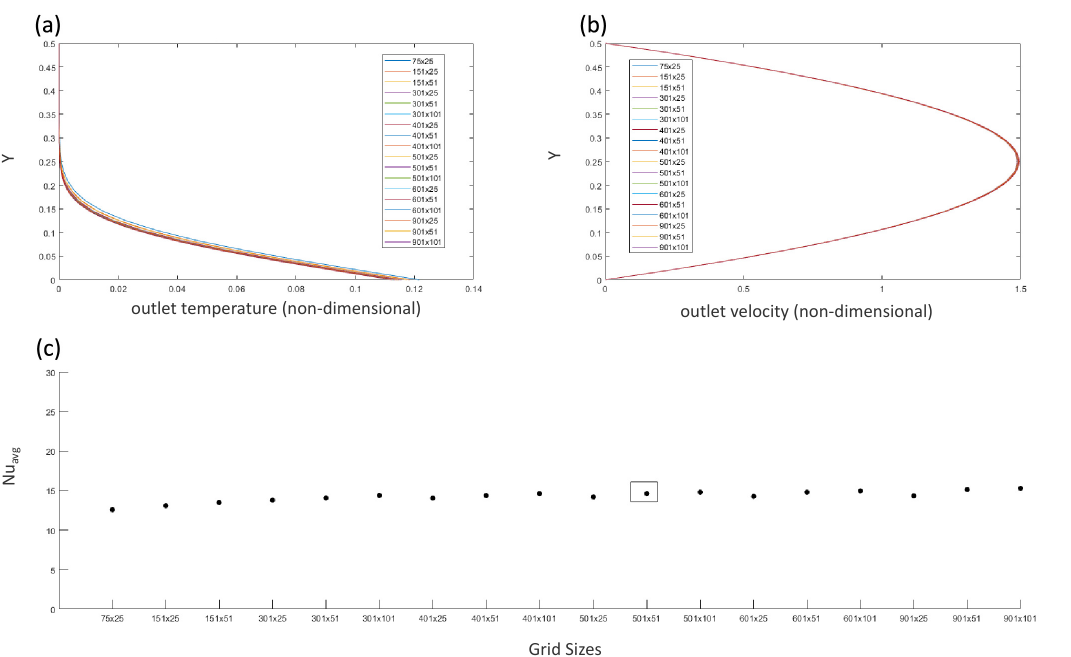}
    \caption{A comparison between (a). the outlet temperature profiles, (b). the outlet velocity profiles and (c) Average Nusselt number of pure water flow through a two dimensional channel at a Reynolds number of 1000 obtained using different grid sizes.}
    \label{Grid1}
\end{figure}
\end{center}

The grid independence of the solver results are checked by simulating the pure water flow through a two dimensional channel with a heated bottom wall. A heat flux of 20 kW/$m^2$ is applied on the bottom wall. A total of 18 different grid sizes were tested, namely, 75x25,  151x25, 151x51, 301x25, 301x51, 301x101, 401x25, 401x51, 401x101, 501x25, 501x51, 501x101, 601x25, 601x51, 601x101, 901x25, 901x51, 901x101. As shown in the Fig.\ref{Grid1}, (a) and (b), all the grid sizes (except for a few coarser grid sizes, see Fig\ref{Grid1}(a)) predicted almost coinciding outlet temperature and velocity profiles. The same deduction is also observed from the average Nusselt numbers obtained using different grid sizes, see Fig.\ref{Grid1}. Instead of choosing either a coarser or a finer grid, a medium grid size of 501x51 is picked from the grid independence test for obtaining further results. The choice of using a medium sized grid with 25551 grid points is to strike a balance between the computational accuracy and computational time.    

\section{Validation against Experimental and Numerical Studies}\label{valid}

\begin{figure}
    \centering
    \includegraphics[scale=0.50]{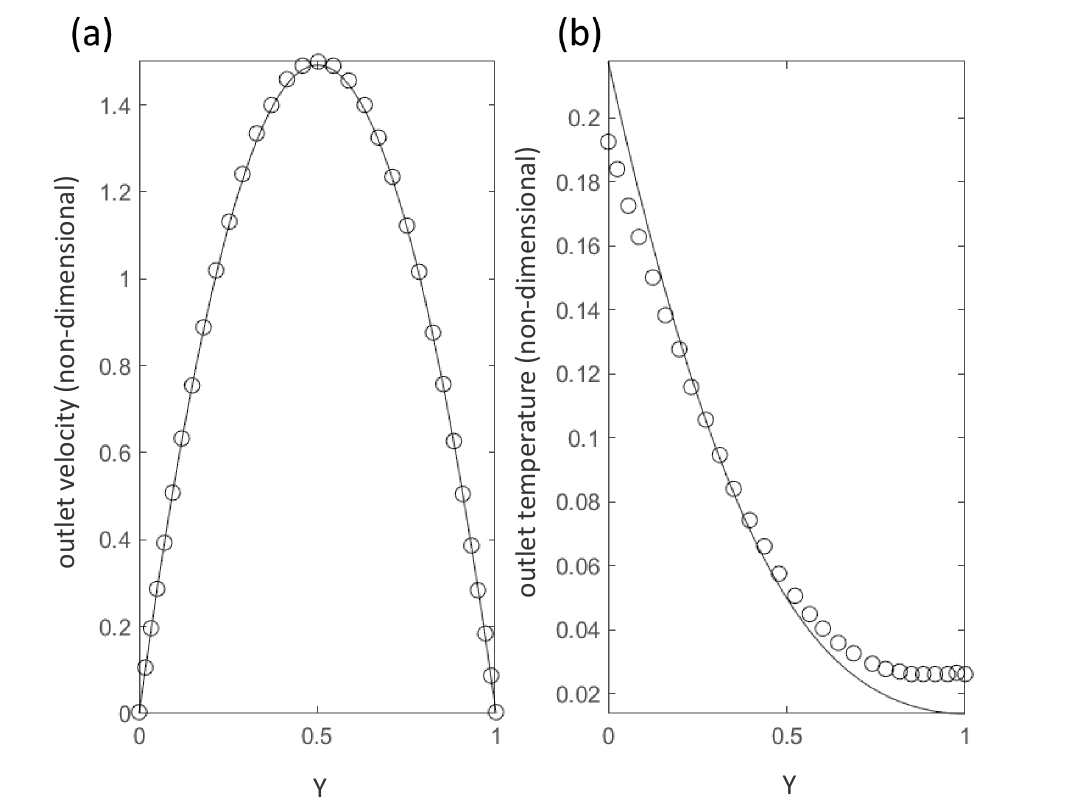}
    \caption{A comparison between (a). the velocity profiles and (b). the temperature profiles of 1\% -Al$_2$O$_3$ -water nanofluid obtained using the current solver and that reported by Kalteh et al. \cite{Kalteh2012} }
    \label{validation1}
\end{figure}

The results of the CFD simulations are obtained using an in-house Fortran-based solver. The solver utilizes a homogeneous model for simulating the nanofluid flows. The results obtained using the model are validated against an experimental and numerical study reported by Kalteh et al. \cite{Kalteh2012}. In their experimental study, Kalteh et al. \cite{Kalteh2012} considered a wide channel with 0.1\% and 0.2\% Al$_2$O$_3$-water nanofluid flow. They also reported a numerical study involving Al$_2$O$_3$-water nanofluid using a homogeneous model. Both the studies involved a channel length of 94.3 mm, and a channel height of 0.580 mm. The channel is heated from bottom using a constant heat flux of 20.5 kW/$m^2$. For validation using the current solver, the same geometry and flow conditions are considered.

\begin{table}[h]
\caption{Comparison of average Nusselt number values with the experimental results of Kalteh et al. \cite{Kalteh2012} \vspace{0.5 cm} }

\label{validation1}

\begin{tabular}{|c|c|cc|c|}
\hline
\multirow{2}{*}{Re} & \multirow{2}{*}{$\epsilon$} & \multicolumn{2}{c|}{$Nu_{avg}$} & \multirow{2}{*}{\% Difference} \\ \cline{3-4}
 &  & \multicolumn{1}{c|}{\begin{tabular}[c]{@{}c@{}}Experimental \\ Study \cite{Kalteh2012} \end{tabular}} & \begin{tabular}[c]{@{}c@{}}Current \\ Solver\end{tabular} &  \\ \hline
225 & 0.1\% & \multicolumn{1}{c|}{7.4} & 7.64 & 3.24 \% \\ \hline
204 & 0.2\% & \multicolumn{1}{c|}{7.5} & 7.52 & 0.27 \% \\ \hline
243 & 0.2\% & \multicolumn{1}{c|}{8.1} & 7.74 & 4.44 \% \\ \hline
\end{tabular}
\label{validationTab1}
\end{table}

The figure \ref{validation1} shows a comparison between the velocity profiles and the temperature profiles of 1\% -Al$_2$O$_3$ - water nanofluid obtained using the current solver and that reported by Kalteh et al. \cite{Kalteh2012}. One can notice from Fig.\ref{validation1} that the velocity profiles obtained using the current solver and that reported by Kalteh et al. \cite{Kalteh2012} completely coincide. It indicates that the current in-house solver is capable of simulating the hydrodynamic behaviour of nanofluid flows in the laminar regime. The temperature profiles do match however there is a slight difference in the profiles. The slight difference in the temperature profiles is due to the fact that Kalteh et al. \cite{Kalteh2012} reported a conjugate heat transfer problem considering some thickness for the bottom wall. The current results obtained using the in-house solver,  neglect the heat conduction through the heated bottom wall. However it is observed from Table \ref{validationTab1} that this assumption does not affect the prediction of the average Nusselt numbers.    

The Table \ref{validationTab1} compares the average Nusselt numbers reported in an experimental study by Kalteh et al. \cite{Kalteh2012} for $0.1\%$ and $0.2\%$ Al$_2$O$_3$-water nanofluid at different Reynolds numbers. As given in the Table \ref{validationTab1}, the maximum difference obtained between the average Nusselt number obtained using the current solver and that reported by Kalteh et al. \cite{Kalteh2012} is only 4.44\%. This result reassures the ability of the in-house solver to predict the thermal behaviour of the nanofluid flows in the laminar regime. Hence this in-house Fortran based solver is used for simulating incompressible - laminar flows involving nanofluids, to study and report the thermal behaviour of nanofluids.

\section{Results}\label{result}

In this study, Al$_2$O$_3$-water nanofluid flows through a wide rectangular channel with a length equal to 25 times the hydraulic diameter (Dh) is considered. The hydraulic diameter (Dh)  is chosen to be twice the height of the channel, which is 0.580 mm, as reported by Kalteh et al. \cite{Kalteh2012}. The two dimensional channel considered here is heated at the bottom wall with a constant heat flux of 20.5 kW/$m^2$, also reported in Kalteh et al. \cite{Kalteh2012}. The nanofluid flow is analyzed at different values of two parameters, namely the Reynolds number, and particle volume concentration. In the subsection \ref{r_ss1}, DMD-based prediction with linear interpolation in one-dimensional parametric space is presented. The parameter considered here is Reynolds number alone. Following that, the DMD-based prediction with bi-linear interpolation in two-dimensional parametric space is presented in the subsection \ref{r_ss2}. There, the parameters considered are Reynolds number and the particle concentration.   

\begin{figure}
    \centering
    \includegraphics[scale=0.5]{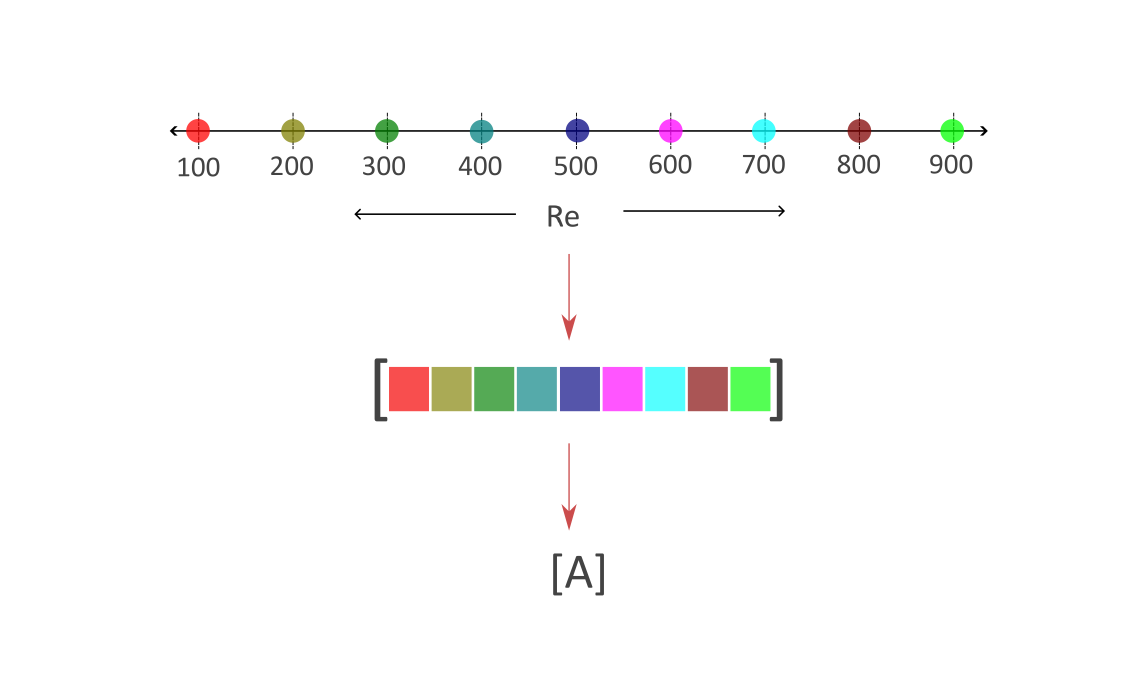}
    \caption{A schematic representation of data instances in the dataset used for obtaining the DMD model in one dimensional parametric space. The stacking of the data instances to obtain the DMD operator is also shown.}
    \label{1D_a}
\end{figure}

\subsection{DMD-based prediction with Linear Interpolation in one-dimensional parametric space}\label{r_ss1}

Firstly, a two dimensional channel with dimensions as mentioned in the section \ref{result}, is considered with a constant heat flux of 20.5 kW/$m^2$, applied uniformly throughout the bottom wall as reported by Kalteh et al. \cite{Kalteh2012}. The nanofluid considered is 1$\%$-Al$_2$O$_3$ - water nanofluid.

\begin{figure}
    \centering
    \includegraphics[scale=0.5]{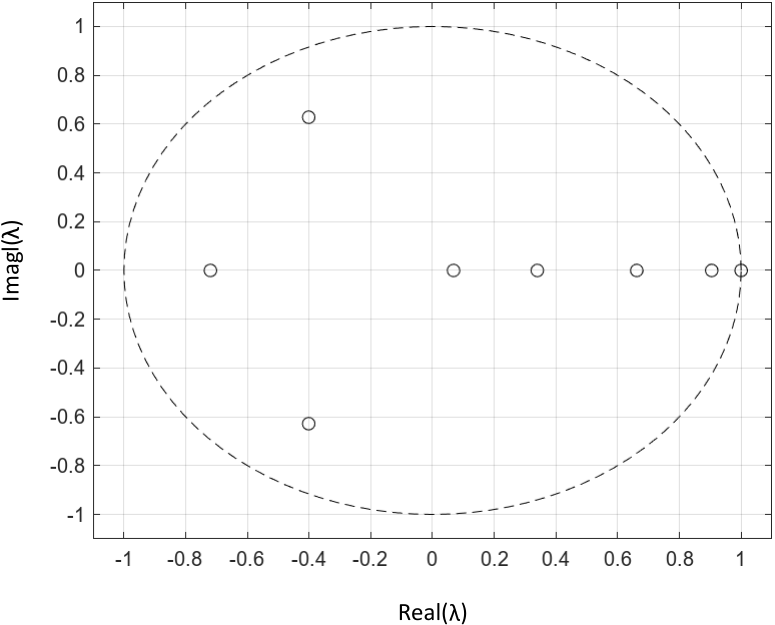}
    \caption{Eigen Values of the DMD operator [A].}
    \label{1D_eig}
\end{figure}

The temperature contours obtained at the quasi steady state at Reynolds numbers Re = 100, 200, .., 900, are used to develop a DMD based model which can be used for making thermal field predictions at new Reynolds numbers. A schematic representation to depict the DMD-based model is given in Fig.\ref{1D_a}. Unlike most of the other studies which make use of DMD for temporal extrapolations, here an attempt is made to used DMD for predictions in the parametric space. Additionally, DMD is used for both future state predictions (extrapolations) and interpolations within the parametric space.  

\begin{figure}
    \centering
    \includegraphics[scale=0.6]{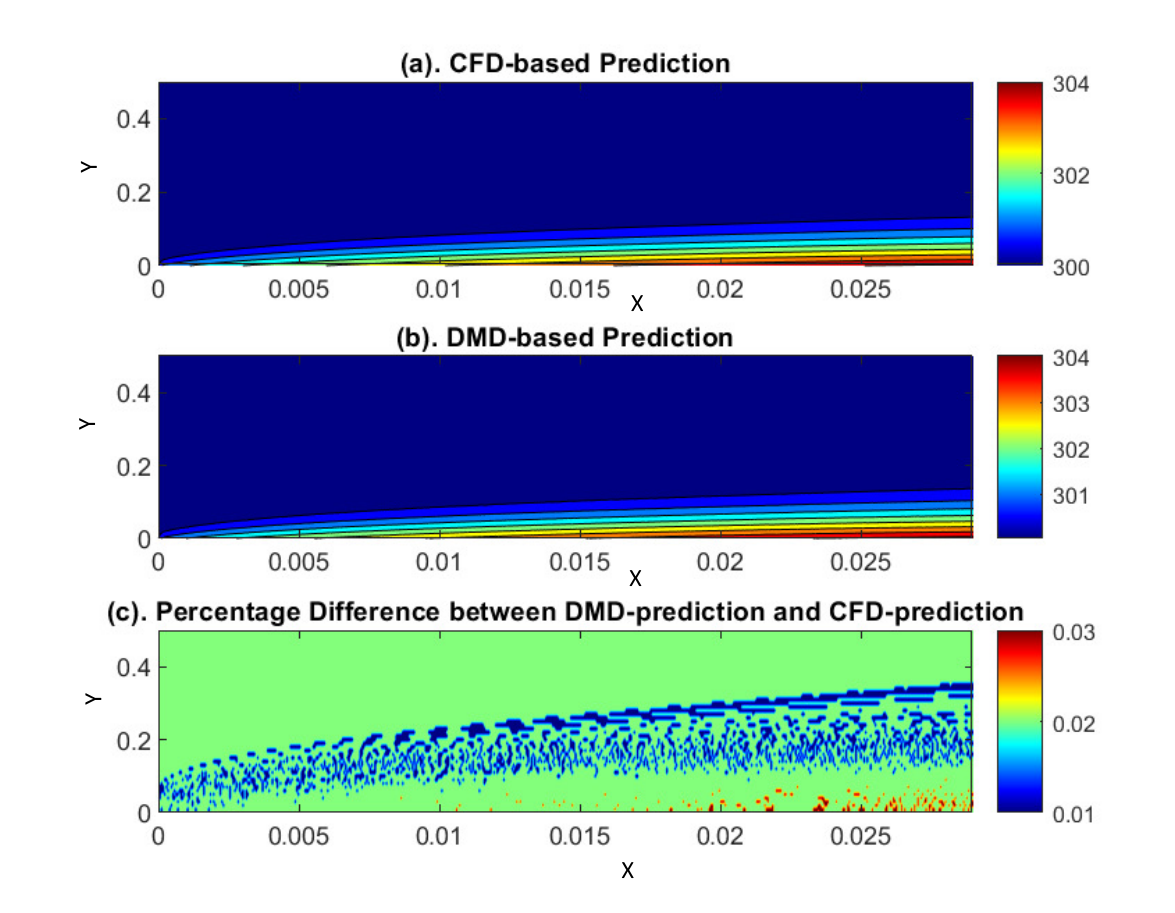}
    \caption{A comparison between the temperature fields predicted using (a). CFD based homogeneous modeling of nanofluid flows, (b). DMD based prediction of the temperature fields at Re = 1000 and (c) the percentage difference between the CFD and DMD based predictions of the temperature fields.}
    \label{1D_tc}
\end{figure}

For developing the DMD based model, the snapshots of the temperature fields at the quasi steady state is collected at Reynolds numbers, Re = 100, 200, 300, .. , upto 900. Using the nine snapshots the DMD operator A is computed. The DMD eigenvalues plotted in the complex plane is given in Fig.\ref{1D_eig}. It is clear from the Fig.\ref{1D_eig} that none of  the identified DMD modes are unstable. This model is used for predictions at various Reynolds numbers and reported in the subsequent subsection \ref{1D_ss1}.   

\subsubsection{Predictions at Re $>$ 1000}{\label{1D_ss1}}

\begin{figure}
    \centering
    \includegraphics[scale=0.6]{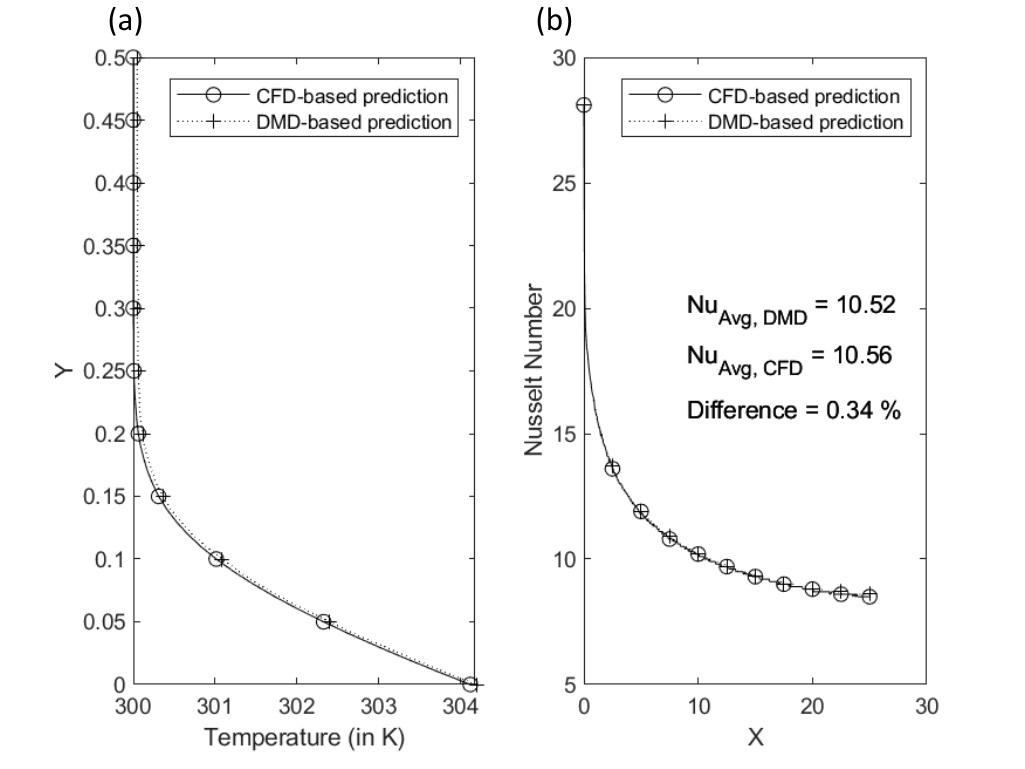}
    \caption{A comparison between (a). outlet temperature profiles and, (b). local Nusselt number along the channel length (non-dimensional), obtained using the CFD and DMD based predictions at Re = 1000.}
    \label{1D_nu}
\end{figure}

The future state prediction in the parametric space considered involves the thermal field prediction at the Reynolds number, Re = 1000 or above. The prediction of the temperature field through the domain obtained using DMD is given in Fig. \ref{1D_tc}, including a comparison with the CFD-based simulation result as well. The maximum percentage difference between the temperature fields predicted using DMD model and that by solving the in-house Fortran based homogeneous solver is only 0.0286\%, while the corresponding mean difference is 0.0170\%. A comparison is also made between the outlet temperature profiles and the local Nusselt number distribution along the channel length, obtained using the DMD-based and the CFD-based predictions. The corresponding results at Re = 1000, is plotted in Fig.\ref{1D_nu}. From Fig.\ref{1D_nu}, one can notice that the outlet temperatures predicted using the CFD based solver and DMD with LI - based prediction are closely matching, an observation expected from the temperature contours plotted in Fig\ref{1D_tc}. The local Nusselt number distribution along the non-dimensionalized channel length, obtained using the approaches coincide too. As mentioned in Fig.\ref{1D_nu}, the difference in the average Nusselt numbers predicted using the two approaches is only 0.34 \%. 

\begin{figure}
    \centering
    \includegraphics[scale=0.5]{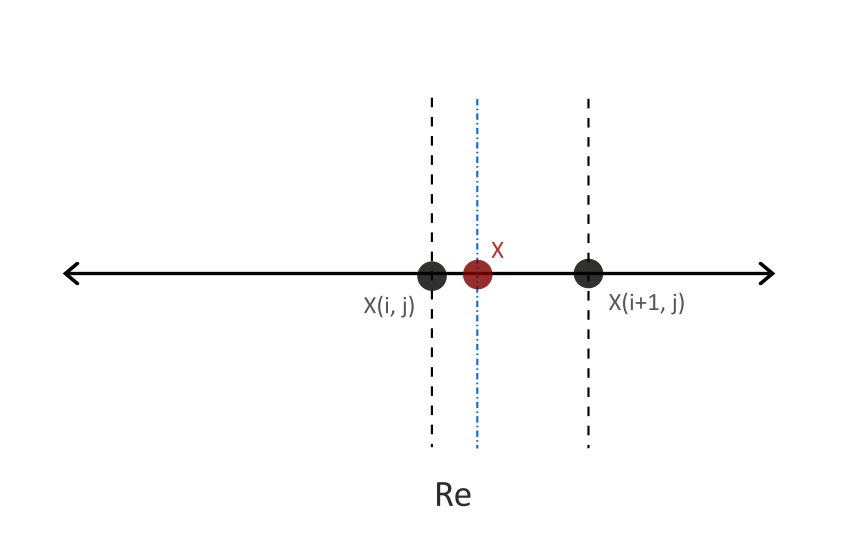}
    \caption{A schematic representation of the linear interpolation used for the prediction of state, X. Note, here the states X(i,j), and X(i+1,j) are computed using the DMD model.}
    \label{1D_b}
\end{figure}

The DMD model is obtained by collecting the thermal fields at regular intervals of Re = 100, 200, $\hdots$, 900. Alternately, the parametric space constituted by the Reynolds numbers is discretized with a step size of 100, within the range [100, 900].   Consequently, the prediction of the thermal field at Reynolds numbers in between the considered intervals, say at Re = 960 cannot be performed directly using the current DMD model. For predicting at such intermediate Reynolds numbers, the concept of linear interpolation is used. A schematic representation of linear interpolation performed is given in Fig.\ref{1D_b}. By employing the linear interpolation, the relationship between the thermal field and the Reynolds number is assumed to be linear within the interval we considered. This is a reasonable assumption to make as well. Hence, for the prediction of temperature fields at Re = 960, a linear interpolation of the temperature fields at Re = 900 and Re = 1000, both predicted using the obtained DMD model, is performed. This assumption of linearity can be exploited for prediction of the thermal field at any Reynolds number outside the regular intervals considered, even within the range of 100 to 900 (see the subsection \ref{1D_ss2}).

\begin{figure}
    \centering
    \includegraphics[scale=0.8]{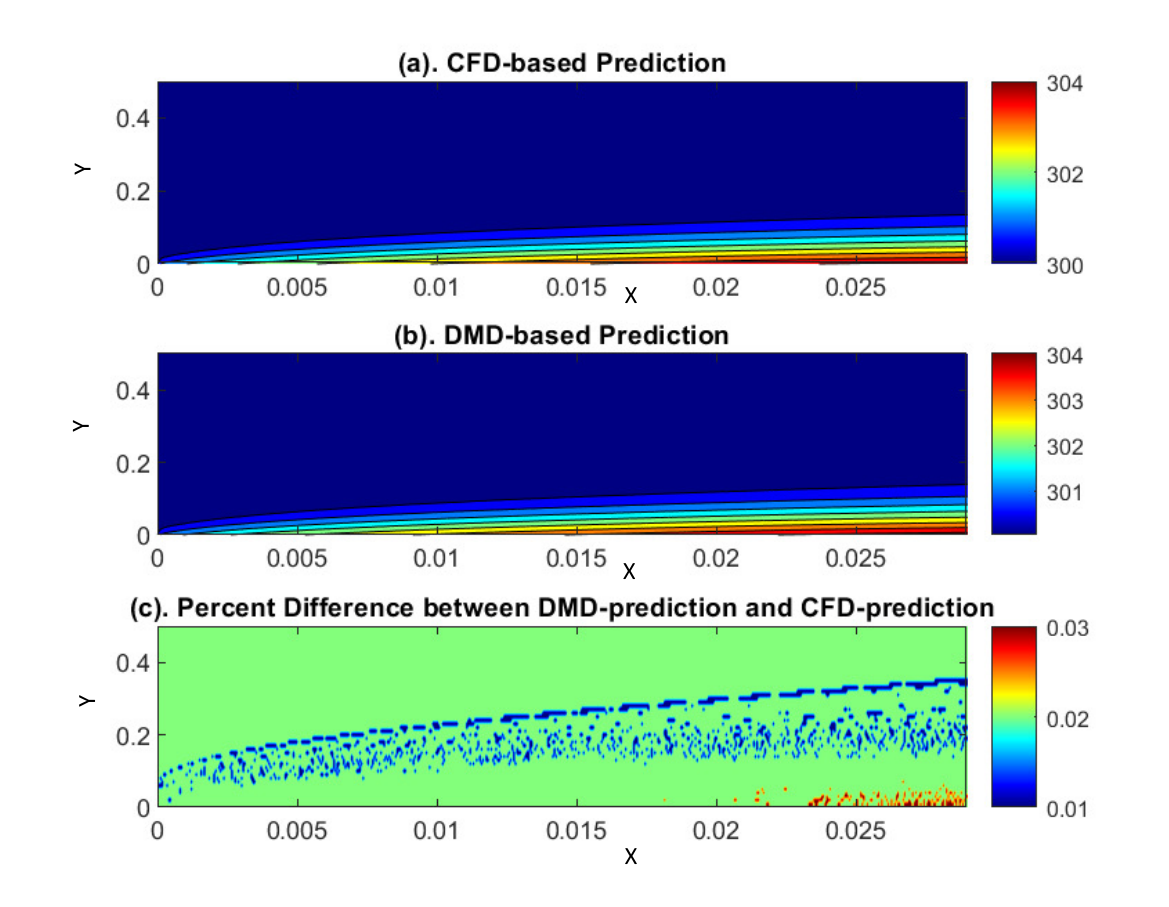}
    \caption{A comparison between the temperature fields predicted using (a). CFD based homogeneous modeling of nanofluid flows, (b). DMD based prediction of the temperature fields at Re = 960 and (c) the percentage difference between the CFD and DMD based predictions of the temperature fields.}
    \label{1D_Re960_tc}
\end{figure}

\begin{figure}
    \centering
    \includegraphics[scale=0.6]{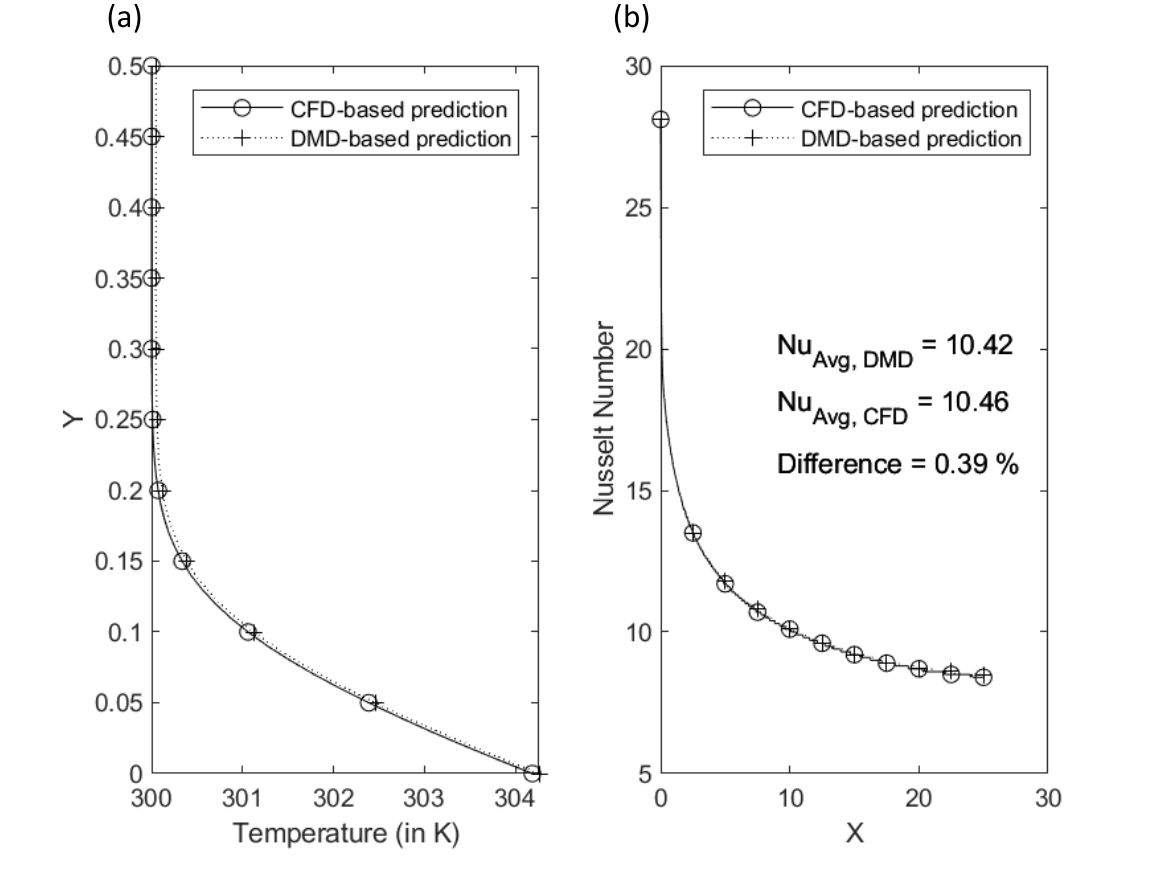}
    \caption{A comparison between (a). outlet temperature profiles and, (b). local Nusselt number along the channel length (non-dimensional), obtained using the CFD and DMD based predictions at Re = 960.}
    \label{1D_Re960_nu}
\end{figure}

The temperature field predicted at Re = 960 is given in Fig.\ref{1D_Re960_tc}. The maximum percentage error in the temperature field prediction using the DMD-LI-based model in comparison with the CFD-based solver is 0.0273\%, while the corresponding mean difference is only 0.0172\%. Additionally, the outlet temperature profiles and the local Nusselt number distribution along the length of the channel obtained using both the methods are compared in Fig.\ref{1D_Re960_nu}. As one can notice from Fig.\ref{1D_Re960_nu}, the difference in the average Nusselt numbers predicted using the both methods is only 0.39\%. In this way, the DMD-based model with linear interpolation (DMD-LI) can be used for prediction of temperature fields at any Reynolds number with considerable accuracy.   

\subsubsection{Predictions using DMD-LI at any Re $>$ 100}{\label{1D_ss2}}

Extending the idea of use of the DMD-LI-based model, an attempt is made here to come up with a data driven model to predict the thermal field of 1\% Al$_2$O$_3$-water nanofluid flow through a 2D channel with a constant wall heat flux of 20.5kW/$m^2$, at different Reynolds numbers $>$ 100. 

\begin{figure}
    \centering
    \includegraphics[scale=0.50]{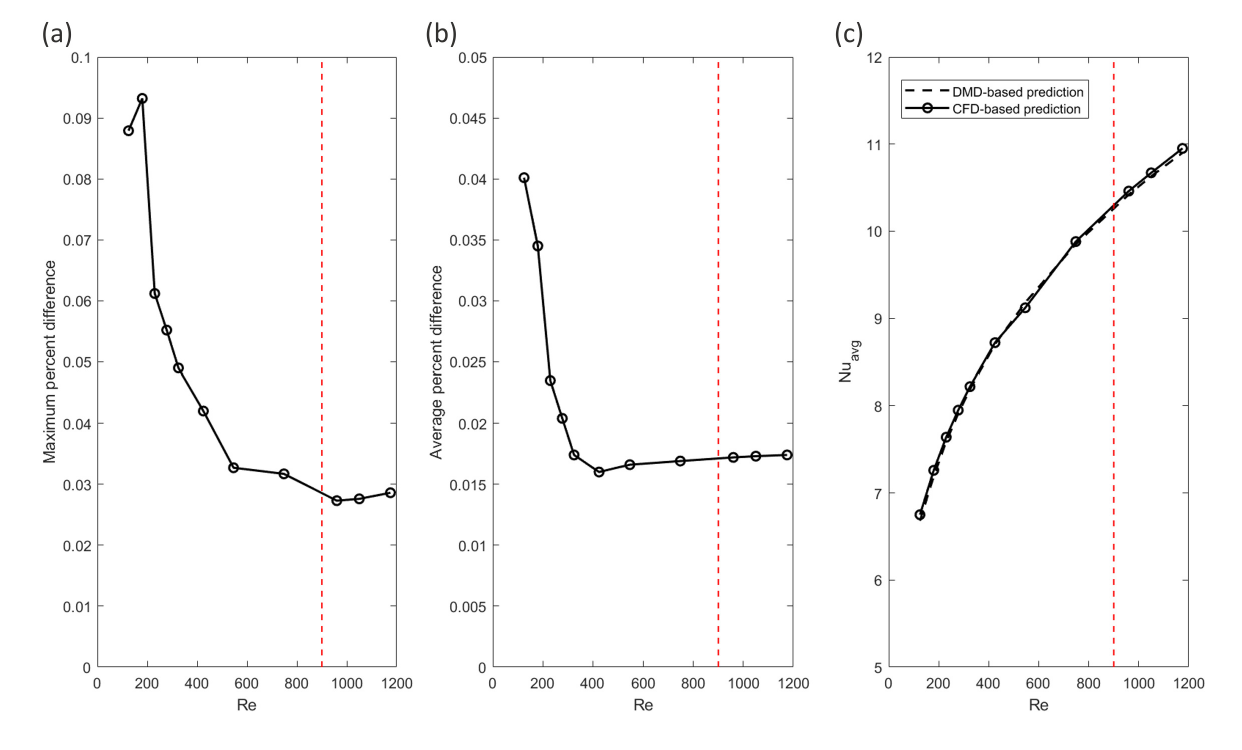}
    \caption{(a) The maximum percentage difference, and (b) the average percentage difference between the CFD-based and DMD-LI-based temperature predictions within the domain under consideration at various Reynolds numbers, and also, (c) the average Nusselt numbers predicted by both approaches at different Reynolds numbers. The vertical red dotted line represents Re = 900.}
    \label{1D_all}
\end{figure}

The details about the prediction are given in Fig \ref{1D_all}. The DMD-LI-based model is used to predict the thermal fields at 11 different Reynolds numbers, namely, 125, 180, 230, 278, 325, 425, 546, 748, 960, 1050, and 1175. The data-driven model is tested on Reynolds numbers both within the training data range of Re=100 to 900 and also outside. As one can notice from Fig \ref{1D_all} (a), the maximum percentage difference between the temperature predictions using the CFD-based model and DMD-LI-based are of the order 10$^{-2}$. Since there is only a very less differences in the predicted temperature fields (see Figs. \ref{1D_all} (a) and \ref{1D_all} (b)), the corresponding average Nusselt number predictions made using the two methods almost coincide with each other, as shown in  Fig.\ref{1D_all} (c).  

To make predictions, the DMD-LI-based model only needs the identified DMD modes, DMD eigenvalues, and the initial dataset (temperature field at Re = 100). As shown here, this proposed data-driven model can predict thermal fields at any Reynolds number within the laminar regime, almost instantly with a commendable accuracy. 

\subsection{DMD-based prediction with Bi-Linear Interpolation in two-dimensional parametric manifold}\label{r_ss2}

In this section, the idea of DMD-LI-based model is extended to a two dimensional parametric space constituted by Reynolds number and particle volume concentration. Here the Reynolds numbers considered are Re = 100, 200, 300, 400, 500, and 600, where as the particle volume concentrations considered include 0.5\%, 0.75\%, 1.0\%, 1.25\% and 1.5\%. Hence a dataset consisting of 30 different data instances are used for identifying the DMD modes and the corresponding DMD eigen values. In this section, all the results are obtained for Al$_2$O$_3$-water nanofluids at different Reynolds numbers and particle volume concentrations. The bottom wall heat flux considered here is 20.5 kW/m$^2$. Here the intention is to develop a DMD-based model with bi-linear (DMD-BLI) model to predict the thermal fields at different Re and $\epsilon$.

\begin{figure}
    \centering
    \includegraphics[scale=0.7]{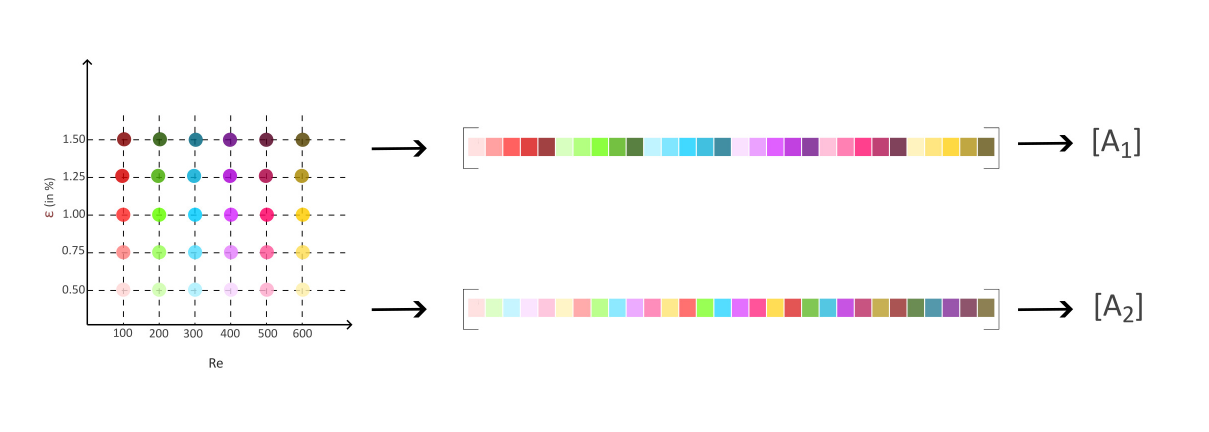}
    \caption{A schematic representation of data instances in the dataset used for obtaining the DMD model in two dimensional parametric space. Two different ways of stacking the data instances to obtain two different DMD operators is also shown.}
    \label{2D_a}
\end{figure}

The dataset in this case, consists of 30 data instances stacked one after the other in a pattern as shown in Fig.\ref{2D_a}. The schematic representation of the data from Fig.\ref{2D_a} also shows the two dimensional parametric space constituted by Reynolds number and the particle volume concentration. Here, there can be two ways of stacking the data instances as given in Fig.\ref{2D_a}. In the first way, the data instances are stacked according to $\epsilon$ followed by Re. In the second way the data instances are stacked according to Re first followed by $\epsilon$. The DMD algorithm applied over each dataset obtained by each way of stacking, yields a different DMD operator. Let the first way of stacking yields [$A_1$] and second one gives [$A_2$]. The difference in the stacking to obtain the two operators  [$A_1$] and [$A_2$] is pictorially depicted in Fig.\ref{2D_a}. 

\begin{figure}
    \centering
    \includegraphics[scale=0.7]{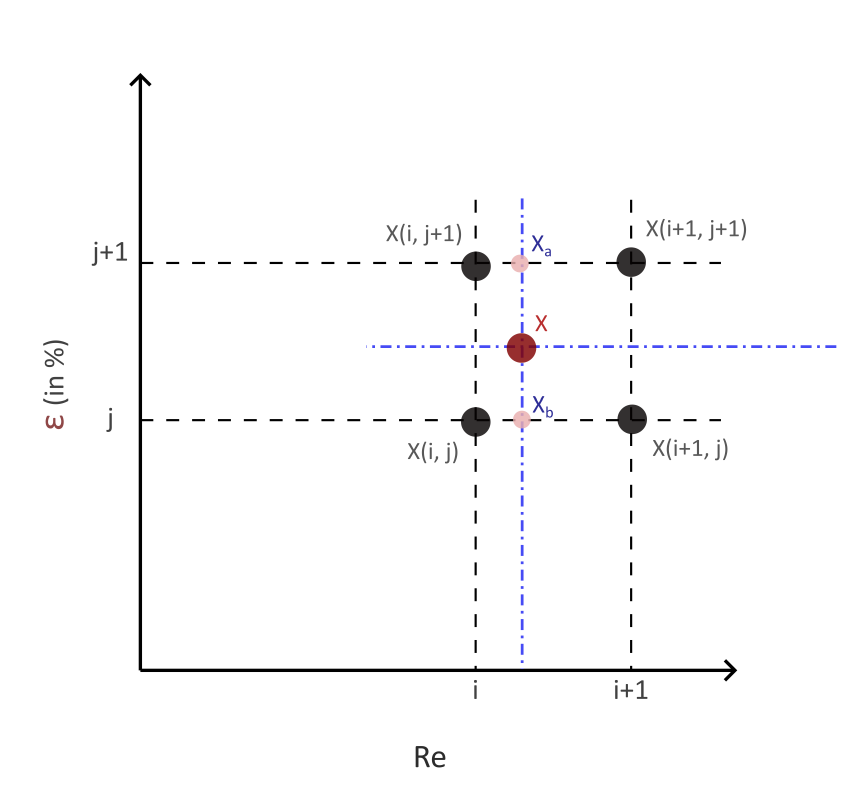}
    \caption{A schematic representation of the bi-linear interpolation used for the prediction of state X. Note, here the states X(i,j), X(i+1,j), X(i,j+1), and X(i+1,j+1) are computed using the DMD model.}
    \label{2D_b}
\end{figure}

The concept of bi-linear interpolation is invoked to predict the temperature fields at any combination of Reynolds number and particle volume concentration. A schematic diagram to represent the concept is given in Fig.\ref{2D_b}. To make the prediction of temperature fields at any combination of Reynolds number and particle volume concentration (represented by point X in Fig. \ref{2D_b}), the prediction of temperature fields at the neighbouring 4 data points (represented by points X(i,j), X(i,j+1), X(i+1, j), and X(i+1,j+1) in Fig.\ref{2D_b}) are used for the interpolation. 

\begin{figure}
    \centering
    \includegraphics[scale=0.55]{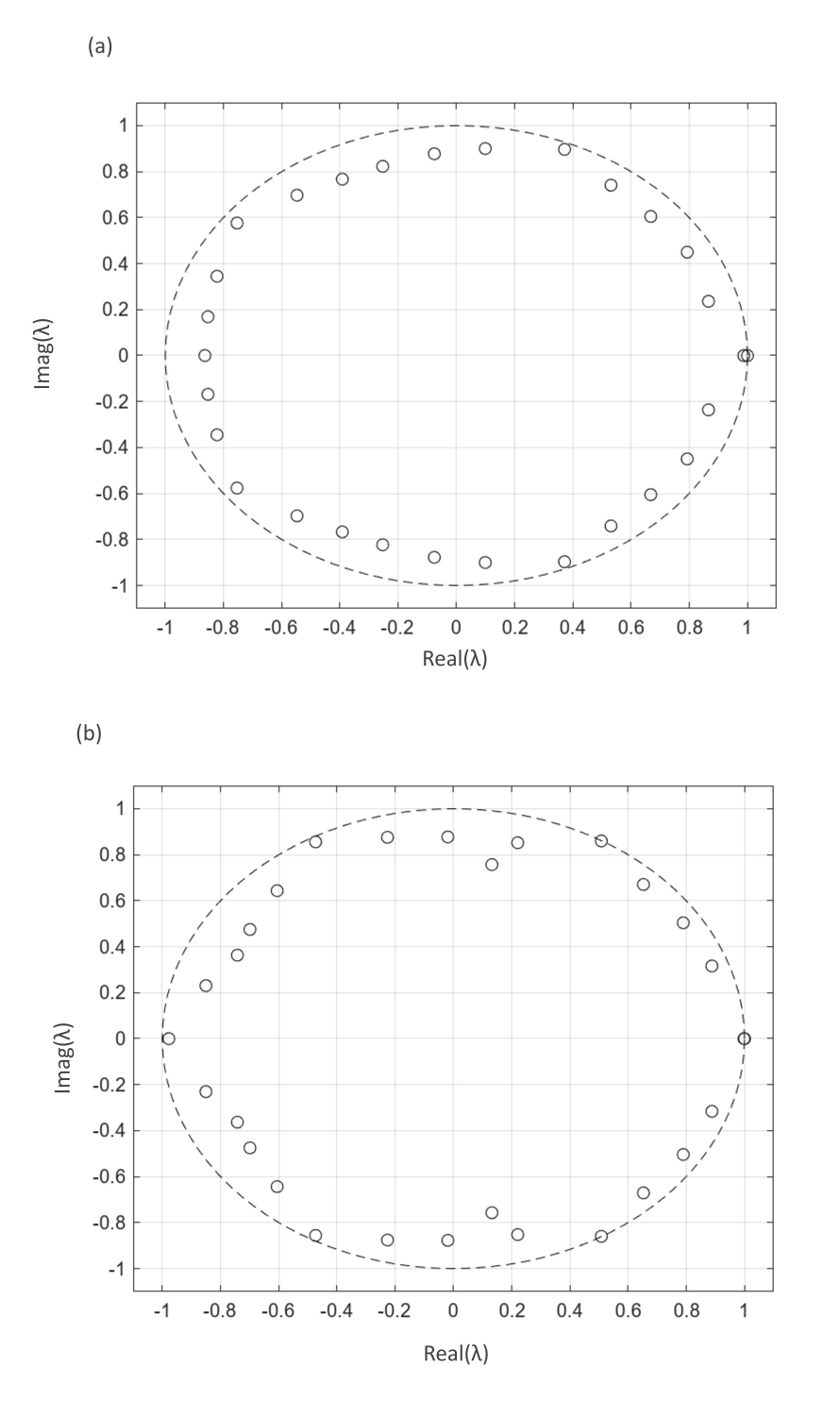}
    \caption{Eigen Values of the DMD operators (a). [$A_1$] and (b). [$A_2$].}
    \label{2D_eigs}
\end{figure}

As already mentioned, two DMD operators are identified by stacking the data instances in two different ways, namely [$A_1$] and [$A_2$]. The DMD eigen values in both the cases are given in Fig.\ref{2D_eigs}. As one can observe from Fig. \ref{2D_eigs}, the DMD eigen values are not outside the unit circle on the complex plane, indicating the absence of unstable dynamics along any of the identified DMD mode directions. Effectiveness of both ways of stacking the data instances in making the predictions are compared in the subsequent subsections (see sub-sections \ref{2D_ss1}, \ref{2D_ss2}, and \ref{2D_ss3}). 

\begin{figure}
    \centering
    \includegraphics[scale=0.6]{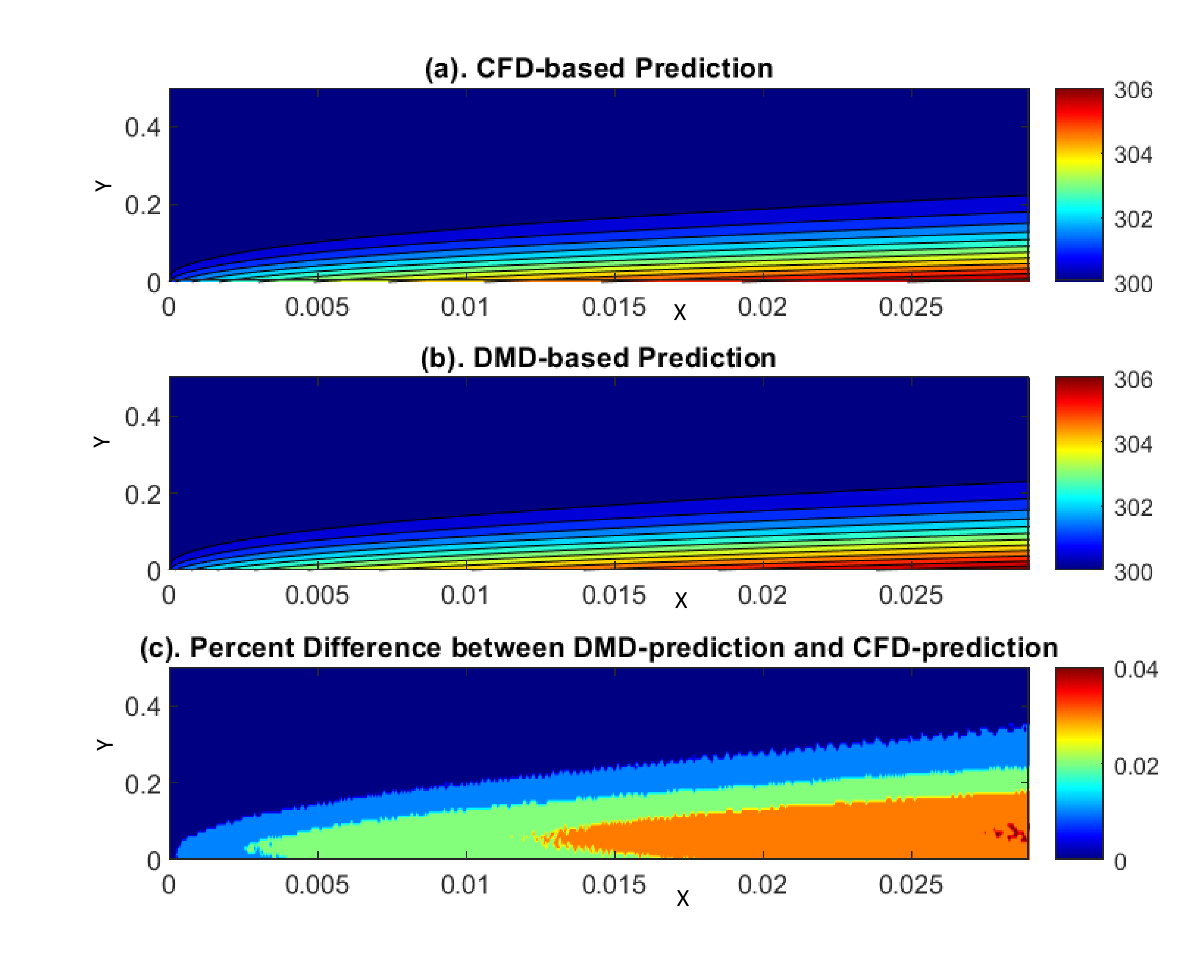}
    \caption{A comparison between the temperature fields predicted using (a). CFD based homogeneous modeling of nanofluid flows, (b). DMD based prediction of the temperature fields using [$A_2$] as the DMD operator, at Re = 325 and $\epsilon$ = 0.065\%, and (c) the percentage difference between the CFD and DMD based predictions of the temperature fields.}
    \label{2D_A1_Re325e0065_tc}
\end{figure}

\subsubsection{ Prediction within 100$\leq$ Re $\leq$ 600 and 0.5\% $\leq$ $\epsilon$ $\leq$ 1.5\%}\label{2D_ss1}

\begin{figure}
    \centering
    \includegraphics[scale=0.6]{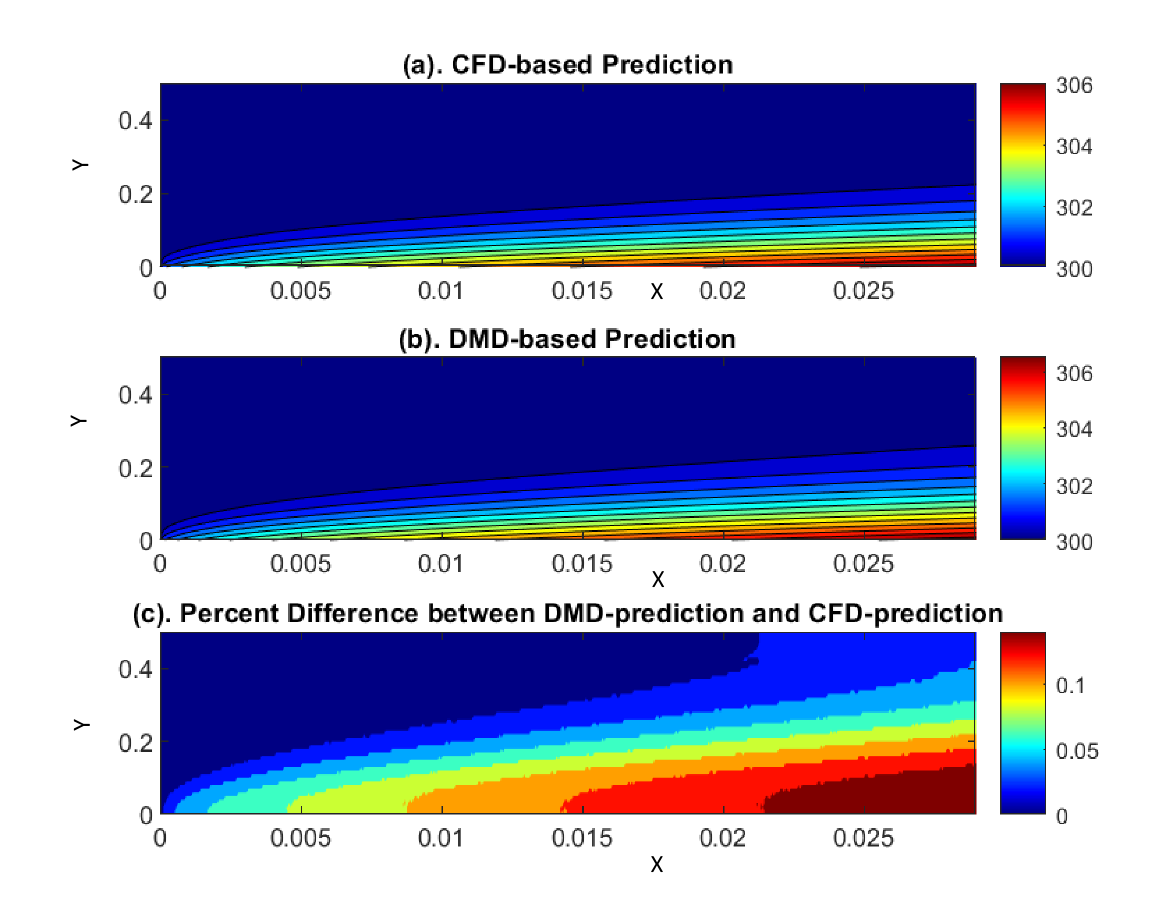}
    \caption{A comparison between the temperature fields predicted using (a). CFD based homogeneous modeling of nanofluid flows, (b). DMD based prediction of the temperature fields using [$A_2$] at Re = 325 and $\epsilon$ = 0.65\%, and (c) the percentage difference between the CFD and DMD based predictions of the temperature fields.}
    \label{2D_A2_Re325e0065_tc}
\end{figure}

\begin{figure}
    \centering
    \includegraphics[scale=0.5]{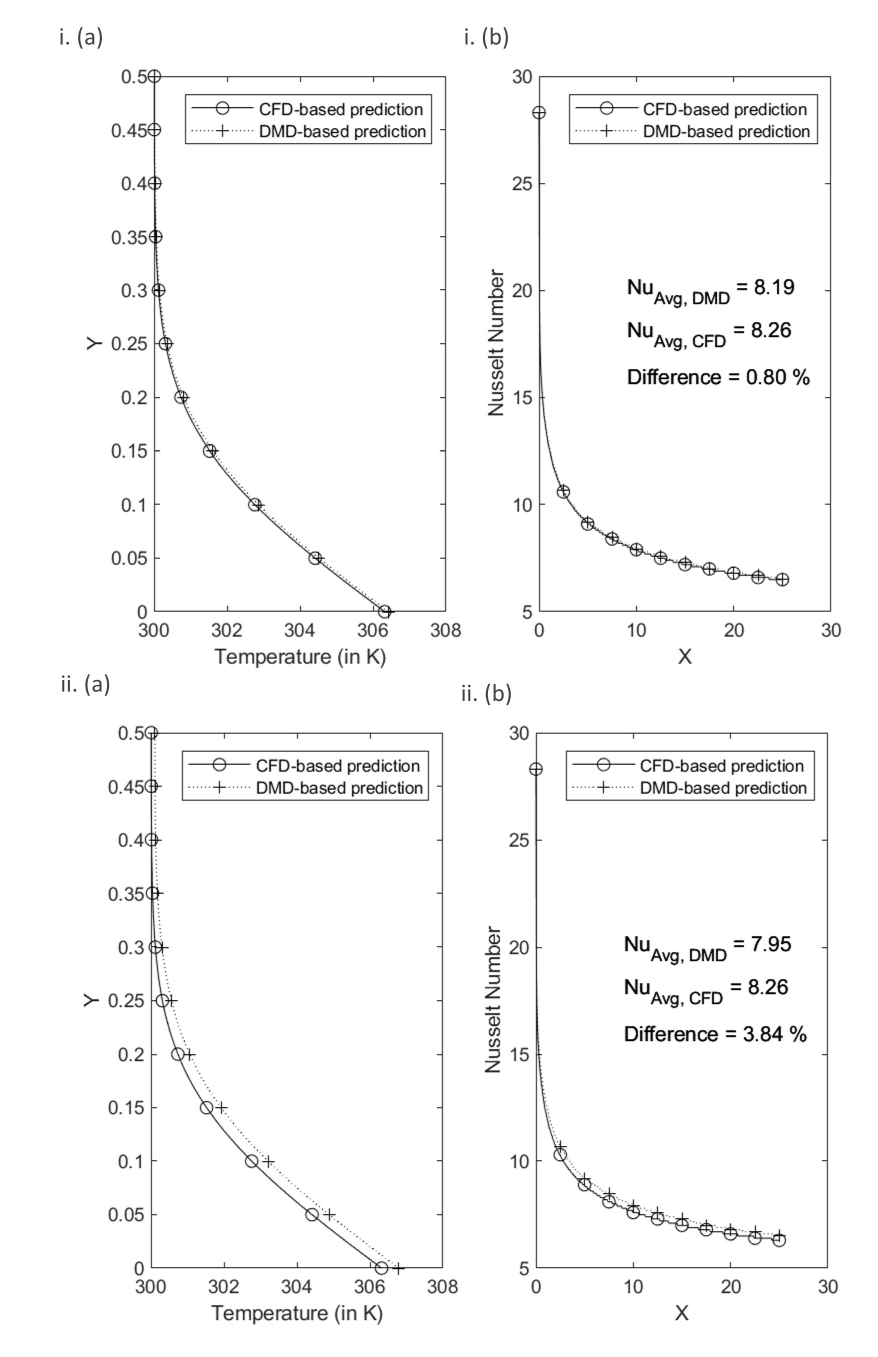}
    \caption{A comparison between (a) outlet temperature profiles and (b) local Nusselt number along the channel length (non-dimensional), obtained using the CFD and DMD based predictions at Re = 325 and $\epsilon$ = 0.65 \%. The subplots `i' and `ii' represent the DMD model using the DMD operators [$A_1$] and [$A_2$] respectively. }
    \label{2D_Re325e0065_nu}
\end{figure}

By making use of the idea of bi-linear interpolation, the temperature field of Al$_2$O$_3$-water nanofluid flow is predicted at Re = 325 and $\epsilon$ = 0.65\%. The predicted temperature fields using the DMD operator [$A_1$] is plotted in Fig. \ref{2D_A1_Re325e0065_tc}, while that of [$A_2$] can be seen in Fig.\ref{2D_A2_Re325e0065_tc}. The temperature contours predicted using the DMD operator [$A_1$] has a maximum percentage difference of 0.036\% while that corresponding to the DMD operator [$A_2$] is 0.155 \%. Consequently the difference in the average Nusselt number predicted using the DMD operator [$A_1$] is only 0.8\%, while that obtained using [$A_2$] is 3.84\%. Overall, the two operators seem to give predictions with acceptable accuracy, however, the DMD operator [$A_1$] outperforms the operator [$A_2$] at Re = 325 and $\epsilon$ = 0.65\%. The same observation is also be made from the comparisons of the outlet temperature profiles and the local Nusselt number distributions given in Fig.\ref{2D_Re325e0065_nu}.

\begin{figure}
    \centering
    \includegraphics[scale=0.70]{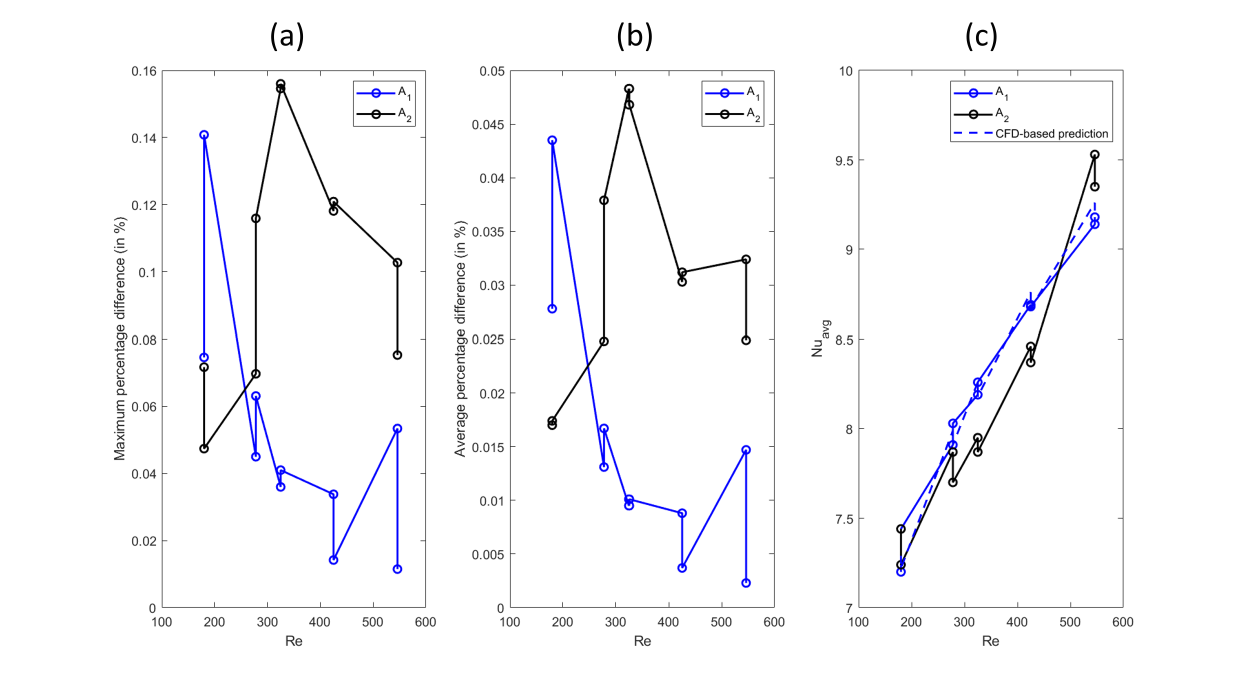}
    \caption{ (a). The maximum percentage difference between the CFD-based and DMD-LI-based temperature predictions using two DMD operators $A_1$ and $A_2$ at various Re and $\epsilon$, (b) the corresponding average percentage differences and also, (c) the average Nusselt numbers predicted by both approaches at different Re and $\epsilon$.}
    \label{2D_all}
\end{figure}

Further, the same two operators are used to predict the temperature fields in ten different cases as given in Fig. \ref{2D_all}. Thermal fields are predicted at a combination of five different Reynolds numbers (180, 278, 325, 425, 546) and two different particle volume concentrations (0.65\% and 1.35\%). It is observed from Fig.\ref{2D_all} that overall both the operators [$A_1$] and [$A_2$] performs well. 
However, the operator [$A_1$] performs well for the predictions at Re $\geq$ 278 and within 0.5\% $\leq$ $\epsilon$ $\leq$ 1.5\% (as seen from the Fig .\ref{2D_all}). Where as, the operator [$A_2$] performs well on the predictions at Re $<$ 278. The maximum percentage difference between the DMD-BLI based and CFD-based predictions of average Nusselt numbers is 2.94 \% for the operator [$A_1$] and 4.03\% for the operator [$A_2$]. The same can be observed from Fig. \ref{2D_all}(c). Since compared to [$A_2$], the DMD operator [$A_1$] predicts the average Nusselt numbers well, [$A_1$] can be considered to be effective for predictions within 100$\leq$ Re $\leq$ 600 and 0.005 $\leq$ $\epsilon$ $\leq$ 0.015.  

\subsubsection{ Prediction at $\epsilon$ $>$ 1.5\%}\label{2D_ss2}

\begin{figure}
    \centering
    \includegraphics[scale=0.8]{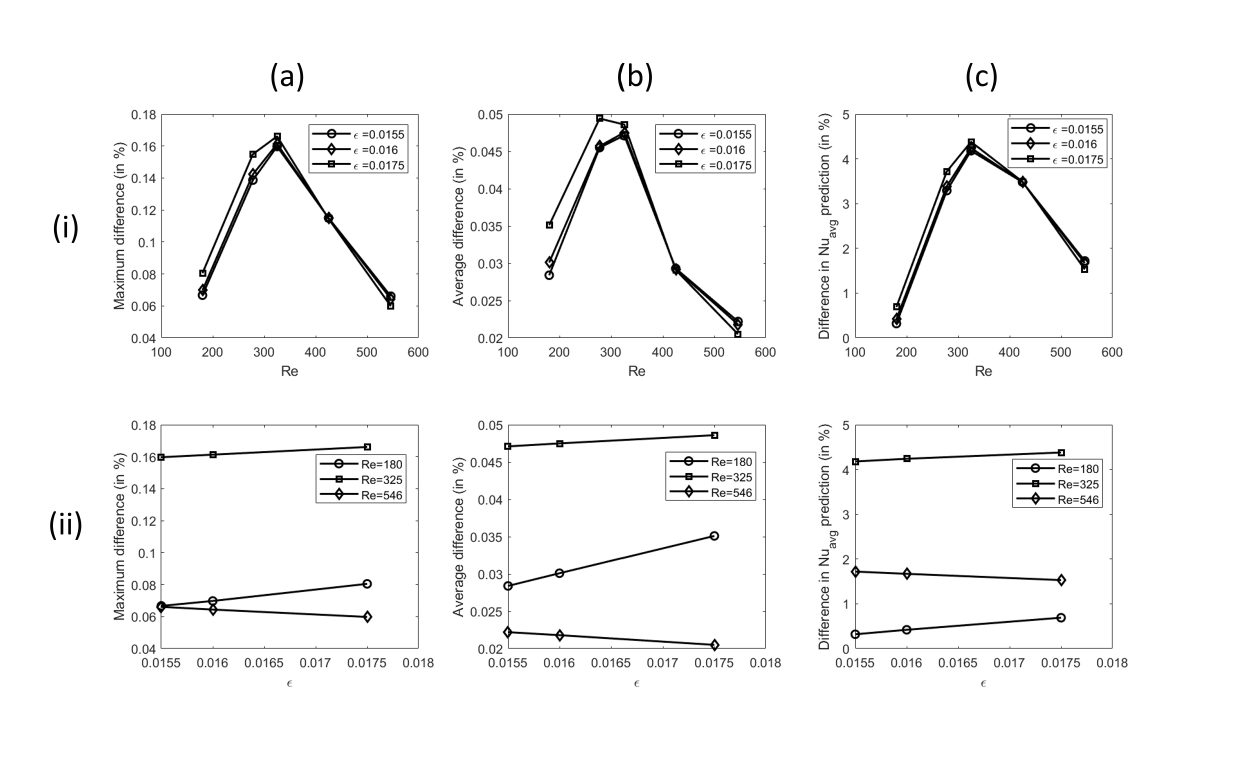}
    \caption{ (a). The maximum percentage difference between the CFD-based and DMD-LI-based temperature predictions various Re, (b) the corresponding average percentage differences and also, (c) the average Nusselt numbers at different Re and $\epsilon$.}
    \label{2D_A2_all}
\end{figure}

The performance of the model for future state prediction is analyzed here. The thermal performance of the nanofluid flow is predicted at concentrations 1.55\%, 1.6\% and 1.75\%, at five different Reynolds numbers, namely, Re = 180, 278, 325, 425, and 546. The Fig.\ref{2D_A2_all} compares the maximum percentage difference in the temperature field prediction, average percentage difference in the temperature field prediction, and percentage difference in the average Nusselt number prediction at different Reynolds numbers and particle concentrations. Firstly, it has been observed that all the three percentage differences are within an acceptable range. The peak maximum percentage difference in the temperature field prediction, observed at Re = 325 and $\epsilon$ = 0.0175, is only 0.167 \%. While the peak value in the case of average percentage difference in the prediction of temperature fields is only 0.0494\%, at Re = 275 and $\epsilon$ = 1.75\%. Whereas, the peak value of the percentage difference in the average Nusselt number prediction, as given in Fig.\ref{2D_A2_all} is only 4.38 \% at Re = 325 and $\epsilon$ = 1.75\%.

As one can notice from Fig.\ref{2D_A2_all} (ii a- ii c), the predictions at $\epsilon$ = 1.55\% are better than at $\epsilon$ = 1.75\%. However, at Re = 546, a contradicting observation is made (see Fig.\ref{2D_A2_all} (ii a- ii c)). Also, it is observed from Fig.\ref{2D_A2_all} (i a - i b) that the maximum and the average percentage errors in the temperature field prediction are lower at Re = 546. This could be attributed to more stable or predictable flow patterns at higher Reynolds numbers, which the model can replicate more accurately.

\begin{figure}
    \centering
    \includegraphics[scale=0.8]{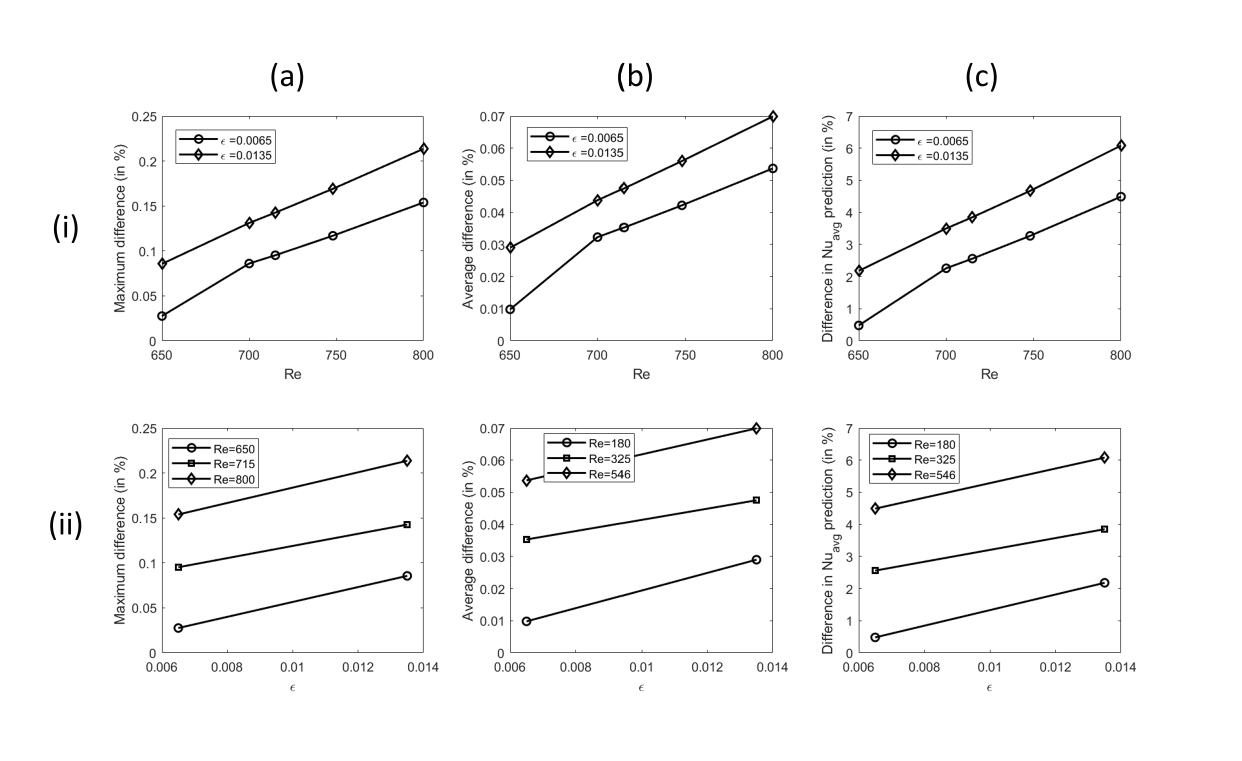}
    \caption{ (a). The maximum percentage difference between the CFD-based and DMD-LI-based temperature predictions various Re, (b) the corresponding average percentage differences and also, (c) the average Nusselt numbers at different Re and $\epsilon$.}
    \label{2D_A1_all}
\end{figure}

\subsubsection{ Prediction at Re $>$ 600}\label{2D_ss3}

The thermal performance of nanofluid flows at Re $>$ 600, namely, Re = 650, 700, 715, 748, and 800, at different particle concentrations $\epsilon$ = 0.65\% and $\epsilon$ = 1.35\% are compared here. Similar to the previous section, the maximum and average percentage differences in the temperature field prediction, along with the percentage difference in the average Nusselt number are compared. 

It has been observed that the peak maximum percentage difference in the temperature field prediction, observed at Re = 800 and $\epsilon$ = 1.35\%, is only 0.214 \%. While the peak value in the case of average percentage difference in the prediction of temperature fields is only around 0.07 \%, also at Re = 800 and $\epsilon$ = 1.35\%. The peak value of the percentage difference in the average Nusselt number prediction, as given in Fig.\ref{2D_A1_all} i(c), is only 6.08 \% at Re = 800 and $\epsilon$ = 1.35\%. For all observations, the performance of the model is better at $\epsilon$ =0.65\%, than that at $\epsilon$ = 1.35\%. The same observation can be made from the Fig. \ref{2D_A1_all}. Similar to the previous case, all the percentage differences between the predictions made using the DMD-BLI-based model and the CFD-based model are well within an acceptable range.   

\section{Conclusions and Future Scope of the Study}

\begin{itemize}

    \item A non-intrusive, parametric DMD-based model is proposed for the prediction of temperature fields of nanofluid flows through a two dimensional rectangular channel.

    \item In this study a DMD based model along with multi-linear interpolation is used for predicting the temperature fields of Al$_2$O$_3$-water nanofluid flows at unseen Reynolds numbers (Re) and particle volume concentrations ($\epsilon$).

    \item The data is obtained using an in-house Fortran based solver. The validation of the homogeneous, laminar, incompressible solver, used for simulating Al$_2$O$_3$-water nanofluid flows through a two dimensional rectangular channel, is also included in this study.
    
    \item Firstly in the DMD with linear interpolation (DMD-LI) based model, the temperature fields collected at 9 different Reynolds numbers are used for identifying the DMD modes and the DMD eigen values. Then those DMD modes and eigen values are used along with linear interpolation for the prediction of temperature fields at any Reynolds numbers greater than 100. 

    \item The DMD-LI based model, predicts temperature fields with a maximum percentage difference of just 0.0273\%, in comparison with the CFD-based solver at Re =960, and $\epsilon$ = 1.0\%. The corresponding difference in the average Nusselt numbers is only 0.39\%. 

    \item The DMD with bi-linear interpolation (DMD-BLI) based model is used for the prediction of temperature field in a parametric space constituted by the Reynolds numbers (Re) and the particle volume concentration ($\epsilon$). 

    \item Two different ways of stacking the data instances yield the DMD operators [$A_1$] and [$A_2$]. Since compared to [$A_2$], the DMD operator [$A_1$] predicts the average Nusselt numbers well, [$A_1$] can be considered to be effective for predictions within 100$\leq$ Re $\leq$ 600 and 0.005 $\leq$ $\epsilon$ $\leq$ 0.015.  

    \item The DMD operator [$A_2$] is used for predictions at $\epsilon$ $>$ 1.5\% and 100$\leq$ Re $\leq$ 600. Similarly, the DMD operator [$A_1$] is used for predictions at Re $>$ 600 and 0.5\% $\leq$ $\epsilon$ $\leq$ 1.5\%. 

    \item When compared to the CFD-based model, the DMD-BLI-based model predicts the temperature fields with a maximum percentage difference of 0.21 \%, at Re = 800 and $\epsilon$ = 1.35\%. And the corresponding percentage difference in the average Nusselt number prediction is only 6.08\%.  

    \item A few of the identified future scope of the current study are as follows:

    \begin{itemize}
    \item A model incorporating more parameters can be developed. However it is observed that addition of more parameters is affecting the accuracy of the predictions. More investigations are required in finding a suitable model which can be used for predictions in multi-dimensional parametric space.

    \item Instead of linear interpolation, use of polynomial and spline interpolations can also be considered for the cases where the assumption of linearity may not be sufficient. 

    \item The model can be extended to include temporal data to predict more complex flows. 
    \end{itemize}
    
\end{itemize}

{ \small

\nomenclature[C]{\(A\)}{DMD operator}
\nomenclature[C]{\(c_p\)}{Specific heat capacity (in J/kgK)}
\nomenclature[C]{\(Dh\)}{Hydraulic Diameter}
\nomenclature[C]{\(n\)}{Time instance}
\nomenclature[C]{\(k\)}{Thermal Conductivity (in W/mK)}
\nomenclature[C]{\(t\)}{Time (in s)}
\nomenclature[C]{\({\bf v}\)}{Velocity Vector}
\nomenclature[C]{\(w\)}{Interpolation weight}
\nomenclature[C]{\({\bf x}\)}{State vector at an instance of time}
\nomenclature[C]{\({\bf X}\)}{Data matrix}
\nomenclature[C]{\({\bf X^\prime}\)}{Time-shifted Data matrix}

\nomenclature[G]{\(\epsilon\)}{Particle volume percentage}
\nomenclature[G]{\(\lambda\)}{Eigen Values of [A]}
\nomenclature[G]{\(\mu\)}{Viscosity (in Ns/m^2)}
\nomenclature[G]{\(\rho\)}{Density (in kg/m^3)}
\nomenclature[G]{\(\theta\)}{Temperature (in K)}
\nomenclature[G]{\(\zeta\)}{Parametric manifold}

\nomenclature[S]{\(nf\)}{Nanofluid}
\nomenclature[S]{\(bf\)}{Base-fluid}

\printnomenclature



\vspace{0.3 cm}

{\flushleft \bf Declaration of generative AI and AI-assisted technologies in the writing process}
\vspace{0.2 cm}

During the preparation of this work the authors used AI-based tools like ChatGPT and MS-Copilot in order to enhance the language quality of the manuscript during its preparation. After using these tools, the authors reviewed and edited the content as needed and take full responsibility for the content of the publication.

\bibliographystyle{unsrtnat}
\bibliography{references.bib}  






\end{document}